# Formation mechanisms and relaxation of NMR spin-echo signals excited by two arbitrary duration radio-frequency pulses in magnets.

M.D Zviadadze[a], G.I. Mamniashvili[a], T.O. Gegechkori[a], A.M. Akhalkatsi[b], T.A. Gavasheli[b],

[a] E.Andronikashvili Institute of Physics of TSU, 6 Tamarashvili st. 0177. Tbilisi, Georgia.

[b] I.Javakhishvili Tbilisi State University, 3 Chavchavadze av. 0128. Tbilisi, Georgia.

m.zviadadze@mail.ru

## Abstract

The work is devoted to the problem of multiple signals of nuclear spin echoes in magnets, excited by a series of radio-frequency (RF) arbitrary duration pulses exceeding the free induction decay time.

The quantum-statistical approach based on the Liouville equation solution for the statistical operator of system is developed for the investigation of echo-processes.

The obtained theoretical results for the number of echo signals, time moments of their formation and their intensities are in good agreement with experiments carried out on magnets (ferrites, ferrometals, half metals, manganites).

The pointed approach is general and could be applied to EPR and NQR, which is interesting also for its application for remote detection of explosives and narcotics.

The application of wide RF pulses and their sequences makes it possible to accumulate weak signals, enriches the echo-response spectrum with clearly separated intensive lines, i.e. essentially increases the sensitivity of apparatus and, correspondingly, the possibilities to explore the fine details of dynamical and relaxation processes taking place in nuclear spin-systems with sufficiently long relaxation times. The work contains also results of new experiments on study of relaxation processes in these systems.

## 1. Introduction

The echo signals in nuclear spin systems (NSS) are usually excited by short pulses of a radio-frequency (RF) field, whose duration $\tau$ is much shorter than the characteristic times of dynamic chaos $T_2^*$. The time $T_2^*$ is associated with so-called inhomogeneous broadening, since this broadening is caused by static inhomogeneities of an object investigated. At $\tau > T_2^*$, there appear new possibilities for the formation of echo signals, which were first noted by Mims [1].

These possibilities were lately studied in many works [2-8] particularly after the observation of single-pulse echo phenomenon (SPE) [9] and explanation of this phenomenon by the non-resonant excitation of NSS of magnets [2]. This mechanism was named as non-resonant one and it is caused by non-adiabatic, jump-like changes of the effective magnetic field in rotating coordinate system (RCS) at edges of RF pulses. In the last years this direction obtained a new development [10-14].

In the present work it is given a short review of quantum-statistical calculation of nuclear spin echo (NSE) signals in magnets excited by two wide RF pulses and presented new experimental results in this direction [13-14].

## 2. Statement of problem

The scheme of the effect of two RF pulses that ensure the formation of multiple echo signals is given in Fig. 1:

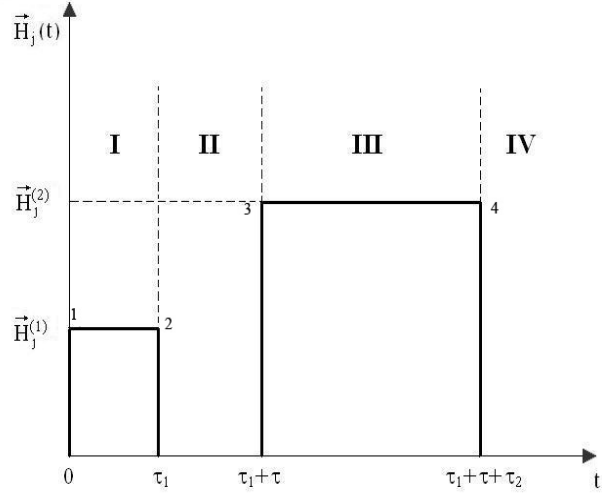

Fig. 1. $\tau_1 < \tau$, $\tau_1 + \tau < \tau_2$. 1, 2, 3, 4 are the pulse fronts.

$$\omega \to \omega^{(1,2)}, \quad \omega_1 = n\gamma_I H_1 \to \omega_1^{(1,2)}, \quad \omega_{nj} = \gamma_I H_{nj} \to \omega_{nj}^{(1,2)}.$$

At the initial time moment $t = 0$ a first RF pulse with the duration $\tau_1$ is applied to the equilibrium NSS. Then, in a time period $\tau > \tau_1$ (time of delay) there occurs a free evolution of the NSS after which at the time moment $t = \tau_1 + \tau$ a second pulse with duration $\tau_2 > \tau_1 + \tau$ is applied. After the end of the second pulse at the time moment $t = \tau_1 + \tau + \tau_2 \equiv t_1$ there again starts a process of free induction, and at times $t > t_1$ there are observed multiple echo signals.

It is assumed that the pulses of the RF field have frequencies $\omega_{nj}^{(1,2)}$, amplitudes $\omega_1^{(1,2)}$ (in the frequency units) and durations $\tau_{1,2}$. It is also assumed that at the time moment at which the pulse is applied to the system there occurs a jump like change in the hyperfine field at the nuclei: $H_{nj} \to H_{nj}^{(1,2)}$ (e.g., due to the displacement of the domain wall caused by a magnetic pulse [11], or for some other reason), which leads to a jump in the resonance frequency of $j$ the isochromate: $\omega_j = \gamma_I H_{nj} \to \omega_j^{(1,2)} = \gamma_I H_{nj}^{(1,2)}$ (it is assumed, as usual, that $H_{nj} \gg H_o$).

## 3. The scheme of calculation of echo signals

In the coordinate system rotating with the frequency of the RF field $\omega$ around $z$ axis (RCS), the NSS is described by a statistical operator $\rho^*(t)$ and by the Liouville equation [12]

$$i\hbar \frac{\partial \rho^*(t)}{\partial t} = [H_t^*, \rho^*(t)], \quad H_t^* = H^* + H_{SL}^*(t), \tag{1}$$

$$H^* = \hbar \sum_j \left(\Delta_j I_j^z + \omega_1 I_j^x\right) + H_{SN}, \quad H_{SL}^*(t) = \exp(i\omega I_z t) H_{SL} \exp(-i\omega I^z t) \tag{2}$$

Here $\Delta_j = \omega_j - \omega$ is the detuning of the resonance, $\omega_j = \gamma_I H_j$ is the Zeeman frequency of the $j$ th isochromate; $\omega_1 = \eta \gamma_I H_1$ is the Rabi frequency (the Rabi inhomogeneous broadening is neglected), $\eta$ – enhancement factor [15], $I_j^x, I_j^y, I_j^z$, $I_j^\pm = I_j^x \pm I_j^y$ are the operators of the nuclear spin $\vec{I}_j$, $H_{SL}$ is the spin-lattice interaction (no its explicit form is required); $H_{SN}$ - is the Suhl-Nakamura interaction.

Below, we neglect the interaction between the nuclear spins belonging to different isochromates; therefore, $H_{SN}$ can be written as follows:

$$H_{SN} \approx \sum_j H_{SN,j}, \quad H_{SN,j} = \sum_{k \neq \ell}^{N_j} U_{k\ell} I_k^+ I_\ell^-, \quad [I_j^z, H_{SN,j}] = 0, \qquad (3)$$

where the summation over $k, \ell$ refers to the spins of the $j$ th isochromate (their number is $N_j$).

As follows from the explicit form of the Hamiltonian $H^*$ in (2), the nuclear spins in the RCS are subjected to an effective magnetic field

$$\vec{H}_j = \left(\Delta_j \vec{k} + \omega_1 \vec{i}\right) / \gamma_I. \qquad (4)$$

The problem consists in solving of the Liouville equation (1) in regions $I - IV$ under the equilibrium initial condition

$$\rho^*(t)\big|_{t=0} = \rho_0 \cong \left(1 - \hbar \beta_L \sum_j \omega_j I_j^z\right) / Tr1, \qquad (5)$$

with the fulfillment of the joining conditions at the time moments $\tau_1$, $\tau_1 + \tau$, $\tau_1 + \tau + \tau_2$:

$$\rho_1^*(0) = \rho_0, \; \rho_{10}^*(\tau_1) = \rho_1^*(\tau_1), \; \rho_2^*(\tau_1 + \tau) = \rho_{10}^*(\tau_1 + \tau), \; \rho_{20}^*(\tau_1 + \tau + \tau_2) = \rho_2^*(\tau_1 + \tau + \tau_2), \qquad (6)$$

which ensure the continuity of the solution. During the action of the RF pulses the spin-lattice interaction is neglected ($H_{SL}^x(t) \ll H_{SN}$).

When solving the Liouville equation in pulsed RF fields, we used the scheme of the calculation described in [12]. The aim of the calculation was the determination of the induction signal defined by the relationship

$$I(t + t_1) = Tr\{\rho^*(t) I^+\} = \sum_j \bar{I}_j^z \exp\left[(i\Delta_j - T_2^{-1})t\right] \cdot \left\{\gamma_j^{(2)} \left[1 - \left(1 - \bar{\gamma}_j^{(1)}\right) \exp(-\tau/T_1)\right] + \right.$$

$$\left. + \alpha_j^{(2)} \gamma_j^{(1)} \exp\left[(i\Delta_j - T_2^{-1})\tau\right] + \beta_j^{(2)} \gamma_j^{(1)*} \exp\left[(-i\Delta_j - T_2^{-1})\tau\right]\right\}. \qquad (7)$$

Explicit expressions for $\alpha_j^{(k)}$, $\beta_j^{(k)}$, $\gamma_j^{(k)}$, $\bar{\gamma}_j^{(1)}$ factors are presented in work [12]. These parameters depend on the angle $\theta_j^{(k)}$ between the axis $z$ and the direction of effective field $\vec{H}_j^{(k)} = \left(\Delta_j^{(k)} \vec{k} + \omega_1^{(k)} \vec{i}\right) / \gamma_I$:

$$\sin \theta_j^{(k)} = \omega_1^{(k)} / \Omega_j^{(k)}, \; \Omega_j^{(k)} = \sqrt{\left[\Delta_j^{(k)}\right]^2 + \left[\omega_1^{(k)}\right]^2}, \; \Delta_j^{(k)} = \omega_j^{(k)} - \omega^{(k)}, \; k = 1, 2.$$

At derivation of expressions (7) it was used following approximations:

$$U^+(\tau_1 + \tau, \tau_1) I_j^\pm U(\tau_1 + \tau, \tau_1) \cong I_j^\pm \exp(-|\tau|/T_2)$$

$$U^+[\tau_1 + \tau, \tau_1] I_j^z U(\tau_1 + \tau, \tau_1) \cong \bar{I}_j^z + (I_j^z - \bar{I}_j^z) \exp(-|\tau|/T_1^{-1}), \qquad (8)$$

where $U[\tau_1 + \tau, \tau_1]$ is the NSS evolution operator from moment $\tau_1$ to moment $\tau_1 + \tau$:

$$U(\tau_1+\tau,\tau_1)=T\exp\left(-\frac{i}{\hbar}\int_{\tau_1}^{\tau_1+\tau}V(t')dt'\right),\ V(t)=H_{SN}+H_{SL}^*(t).$$

Approximations (8) provide the equivalency of quantum approach to the solutions of Bloch equations in regions of free evolutions $II, IV$.

## 4. Classification of echo signals

In the general case the quantity (7) contains 20 signals, namely, two induction signals corresponding to the moments $t=\tau_1$ and $t=t_1$, and 18 echo signals, which can be classified as follows [13,17]:

1. Six primary TPE signals (**12**), (**13**), (**14**), (**23**), (**24**), (**34**), which are formed pairwise by the fronts 1, 2, 3, 4 ((**12**) and (**34**) are signals of the single-pulse echo (SPE) from the first and second RF pulses, respectively).

2. Four secondary TPE signals ((**12**)**3**), ((**12**)**4**), ((**13**)**4**), and ((**23**)**4**), which are formed by the primary echo signal with the subsequent fronts 3, 4.

3. Four signals of primary stimulated echoes (**123**), (**124**), (**134**), and (**234**), which are formed by triples of fronts: the longitudinal component of nuclear magnetization, created by signals (**12**), (**13**), (**23**) and reading fronts 3, 4. At $\Omega_j^{(2)}\tau_2<\Omega_j^{(1)}\tau_1$ for $t>t_1$ time it is observed only two signals of stimulated echo (**123**), (**124**) [14]: at $\Omega_j^{(2)}\tau_2>\Omega_j^{(1)}\tau_1$ it is observed three signals of stimulated echo (**124**), (**134**), (**234**).

4. Two signals of secondary TPE(s) (((**12**)**3**)**4**), ((**123**)**4**), which are formed by the echo signal ((**12**)**3**) and by the signal of the stimulated echo (**123**) with the front 4, correspondingly.

5. One echo signal of a "complex" stimulated echo ((**12**)**34**) formed by three elements: by the signal (**12**) and fronts 3 and 4.

6. One signal (**1234**) is formed by longitudinal component of nuclear magnetization $m_z$ created by the first pulse, and by the second pulse (by the fronts 3, 4): allowing for the phase $i\Omega_j^{(1)}\tau_1$ accumulation during $\tau_1$ in $m_z$ which doesn't change in the interval $[\tau_1,\tau_1+\tau]$. Then due to the second pulse there is additional accumulated phase $i\Omega_j^{(2)}\tau_2$, and, therefore, signal (**1234**) is observed at the moment $t\approx\tau_1+\tau_2$ [*)].

At any relation $\Omega_j^{(2)}\tau_2>\Omega_j^{(1)}\tau_1$ or $\Omega_j^{(2)}\tau_2<\Omega_j^{(1)}\tau_1$, the maximum number of echo signals that can be observed after the end of the second pulse is equal, just as in [7, 17], to thirteen (in the first case it is not observed signals (**12**), ((**12**)**3**), (**123**), (**23**), (**13**), and in the second case it is absent signals (**12**), (**134**), (**234**), (((**12**)**3**)**4**), ((**123**)**4**)), whose moments of the appearance lie to the left of the moment of the end of the second pulse $t_1$). Analogically, in the case $\tau_2<\tau_1<\tau$ the signals (**12**), (**134**), (**234**), (((**12**)**3**)**4**), ((**123**)**4**) are not observed. The real number of observed signals and the moments of their appearance depend on the relationship between $\tau_1,\tau,\tau_2$ and on the magnitudes of jumps.

## 5. Particular cases

The calculation gives (in some cases it is presented the explicit expressions of signals):

1. $\tau_2=\tau=0$. It is easy to show, that the factors $\gamma_j^{(2)}=\beta_j^{(2)}=0$, $\alpha_j^{(2)}=1$ and the relationship (7)

---

[*)] In work [17] the formation mechanism of the (**1234**) signal is not explain.

is transformed into the free induction signals from the first pulse:

$$I(t+t_1) = \sum_j \bar{I}_j^z \exp\{(i\Delta_j - T_2^{-1})t\}\gamma_j^{(1)} \qquad (9)$$

2. $\tau \gg T_1$. The relationship (7) is transformed into the free induction signal from the second pulse:

$$I(t+t_1) = \sum_j \bar{I}_j^z \exp[(i\Delta_j - T_2^{-1})t]\gamma_j^{(2)} \qquad (10)$$

3. $\tau = 0$. In this case it is obtained the result of work [12]:

$$I(t+t_1) = \sum_j \bar{I}_j^z \exp[(i\Delta_j - T_2^{-1})t](\bar{\gamma}_j^{(1)}\gamma_j^{(2)} + \gamma_j^{(1)}\alpha_j^{(2)} + \gamma_j^{(1)*}\beta_j^{(2)}) \qquad (11)$$

At time $t > \tau_2 > \tau_1$ it is observed four echo signals: (**23**) from fronts 2 and 3 at moment $t_{2e} \approx \tau_2$, (**13**) from fronts 1 and 3 at moment $t_{4e} \approx \tau_1 + \tau_2$, ((**12**)**3**) from echo signal (**12**) and front 3 at moment $t_{3e} \approx \tau_2 - \tau_1$ and one signal of stimulated echo (**123**) at moment $t_{1e} \approx \tau_1$, decaying according to the law of $\exp(-\tau_1/T_2^{(1)})$ when it is considered that $T_1^{-1} \approx 0$. It is the consequence of the fact that in the strong RF fields ($\omega_1 \gg T_2^{-1}$) relaxation times $T_1$ and $T_2$ are mixed both in quantum and classical [16, 18] approaches.

4. $\tau \gg T_2^{-1}$. The transverse components of nuclear magnetization, created by the first pulse, disappear. Only the longitudinal component remains which reduced the free induction signal from the second pulse:

$$I(t+t_1) = \sum_j \bar{I}_j^z \exp[(i\Delta_j - T_2^{-1})t] \cdot [1-(1-\bar{\gamma}_j^{(1)})\exp(-\tau/T_1)]\gamma_j^{(2)} \qquad (12)$$

The signal (12) is increased with the increase $\tau$ and attain the maximum (10) at $\tau \gg T_1$. In general case for $t > \tau_2$ time (12) contains four echo signals.

$$(\mathbf{34}) = \sum_j \bar{I}_j^z \sin\theta_j^{(2)} \sin^2(\theta_j^{(2)}/2)(1-\sin^2\theta_j^{(1)}\exp(-\tau/T_1))\exp(-t/T_2 - \tau_2/T_{2j}^{(2)}), \quad t \approx \tau_2; \qquad (13)$$

Obviously, the expression $1-\sin^2\theta_j^{(1)}\exp(-\tau/T_1) > 0$ and increases with the increase of $\tau$. That is the allowing for the influence of the first pulse decreases the SPE amplitude from second pulse.

$$(\mathbf{1234}) = \frac{1}{2}\sum_j \bar{I}_j^z \sin^2\theta_j^{(1)} \sin\theta_j^{(2)} \sin^2\frac{\theta_j^{(2)}}{2}\exp\left(-\frac{t}{T_2} - \frac{\tau}{T_1} - \frac{\tau_1+\tau_2}{\bar{T}_{2j}}\right), \quad t \approx \tau_1 + \tau_2, \qquad (14)$$

$$\frac{1}{\bar{T}_{2j}} = \frac{1}{\tau_1+\tau_2}\left(\frac{\tau_1}{T_{2j}^{(1)}} + \frac{\tau_2}{T_{2j}^{(2)}}\right);$$

$$(\mathbf{124}) = \sum_j \bar{I}_j^z \sin^2\theta_j^{(1)} \sin\theta_j^{(2)} \cos\frac{\theta_j^{(2)}}{2}\exp\left(-\frac{t}{T_2} - \frac{\tau}{T_1} - \frac{\tau_1+\tau_2}{\bar{T}_{2j}}\right), \quad t \approx \tau_1; \qquad (15)$$

$$(\mathbf{123}) = -\frac{1}{2}\sum_j \bar{I}_j^z \sin^2\theta_j^{(1)} \sin\theta_j^{(2)} \cos^2\frac{\theta_j^{(2)}}{2}\exp\left(-\frac{t}{T_2} - \frac{\tau}{T_1} - \frac{\tau_1+\tau_2}{\bar{T}_{2j}}\right), \quad t \approx \tau_1 - \tau_2. \qquad (16)$$

The value $\bar{T}_{2j}$ is the effective time of transverse relaxation of signals (**123**), (**124**), (**1234**) due to the excitation by two long RF pulses. When the jumps are absent then $T_{2j}^{(1)} = T_{2j}^{(2)} = T_{2j}$ and $\overline{T}_{2j} = 2T_{2j}/2$, i.e. twice shorter than the SPE transverse relaxation time. In general case $\overline{T}_{2j}$ is intermediate between $T_{2j}^{(1)}$ and $T_{2j}^{(2)}$.

## 6. Results and discussions

The approach used in the theoretical part of work is the quantum-statistical generalization of approaches developed in [2, 7] using, correspondingly, classical Bloch equations and statistical tensors method. In work [2] so-called non-resonant mechanism of SPE formation was presented for the first time and in [7] – multiple-pulse mechanism of SPE formation allowing for possible non-equilibrium of spin system before the consecutive PF pulses in the RF pulse train.

As it was shown in our works [19] the best magnet to compare theoretical predictions of this work to experimental ones is lithium ferrite, because in this magnet it is absent echo signals formed by the distortion mechanism [6] appearing when front of RF pulses are distorted due to the transition processes in the RF mains.

The experiments were carried out by the incoherent spin-echo spectrometer [6] at liquid nitrogen temperature.

Circular disks of diamagnetically diluted ferrites $Li_0Fe_{25-x}Zn_xO_4$ ($0 \leq x \leq 0.25$) of 12-15 mm in diameter and 5-8 g in weight enriched by $^{57}Fe$ iron isotope (96.8%) were used. Besides it was used improved resonator system of spectrometer as in [19] and signals a nuclear spin-echo were averaged using a Tektronix 2430 Å digital memory spectrometer.

We use approach of work [12] where it was studied the multiple-pulse analogs of SPE obtained upon jump-like changes of the effective magnetic field direction in the RCS within pulse duration. It turned out that the pulse fronts and the position of jump-like change in the direction of $\vec{H}_j^{(k)}$ in an RCS within the RF pulse have a quantitative analog-exciting RF pulses in the Hahn echo method. In this case, the magnitude of the change in the direction of $\vec{H}_j^{(k)}$ in an RCS is an analog of the angle of deviation of the vector of nuclear magnetization under the effect of RF pulses in the Hahn method.

In frames of this approach, the action of two wide RF pulses can be considered as being equivalent to the action of a single pulse, when apart from the pulse fronts within the duration of RF pulse there are occur two jump-like changes of $\vec{H}_j^{(k)}$ in the RCS, and the amplitude of the RF field between them is zero. Such a complex single-pulse action is analogous to a four-pulse action in the Hanh method [12].

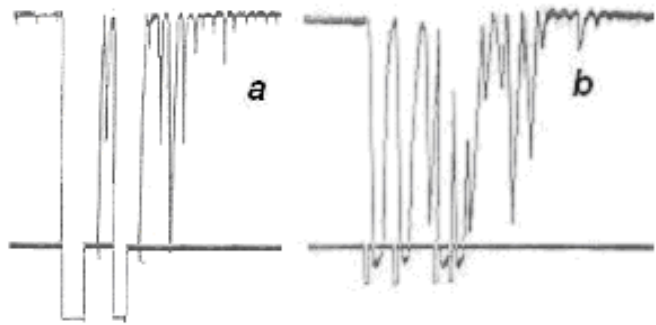

Fig. 2. Oscillograms of the signals of multiple echo upon the excitation by two wide RF pulses in lithium ferrite: (a) at $f_{NMR} = 71$ MHz, lengths of the RF pulses $\tau_1 = 8$ μs, $\tau_2 = 5$ μs, and the time span between them $\tau_{12} = 9$ μs; and (b) upon the action of short ($\tau_p = 1$ μs) RF pulses coinciding with the edges of wide pulses.

Fig. 2 displays an oscillogram of multiple echo signal upon the action of two pulses of different duration and is similar to that one obtained at action of four short RF pulses coincide with the edges of two wide RF pulses [13].

The period of repetition of a pair RF pulses was optimum for the observation of multiple echo. The upper beam shows the amplitude of the echo signals from the NMR receiver depending on the time; the lower beam depicts the signals from video-detector, illustrating the shape amplitude, and duration of the RF pulses. The signals of multiple echo have 13 components in the correspondence with theoretical results and experimental results obtained for multiple-echoes in works [7, 17]. The high intensity of observed multiple echo signals in lithium ferrite made it possible to establish the nature of fast relaxing component in multiple-echo signal [10] for the first time observed in [4] on the example of $FeV$.

The existence of such short relaxation times of one of the main components upon the increase in the duration of the RF pulse can be understood taking into account the fact (firstly noted in [2]) that the SPE signal in terms of the non-resonant mechanism can have a relaxation time shorter than $T_2$, since the conditions of dephasing of the isochromates in the effective magnetic field $\vec{H}_j^{(k)}$ in the presence of an RF pulse differ from those observed in the process of their dephasing after the end of the action of the RF pulse. In such magnets as $Co$ and $Co_2MnSi$ it is effective both non-resonant and distortion mechanisms of SPE formation allowing the observation of strongly relaxing component similar to [4] and component formed by the distortion mechanism with longer transverse relaxation time [14].

### 7. Conclusions

The theoretical and experimental study of multiple echo-signals from two wide RF pulses was carried out. The experimental results coincide with the theoretical predictions. These data are useful for using methods of echo spectroscopy for studying both general problem of dynamic chaos and local characteristics of magnetic materials.

The use of wide pulses opens new possibilities to study different signals of multiple echoes differently connected with the properties of studied magnetic materials.

**Acknowledgements:** This work is supported by the Shota Rustaveli National Science Foundation, Short-Term Individual Travel Grant 2012_tr_252. The authors are grateful to L. Zamtaradze and M. Nikoladze for help with processing articles

**REFERENCES**

1. W.B. Mims. Phys. Rev. **141** (2), 499–502 (1966).

2. V.P. Chekmarev, M.I. Kurkin, and S.I. Goloshchapov. ZhETF, **76** (5), 1675–1684 (1979). (Russian).

3. G.I. Mamniashvili and V.P. Chekmarev. Bullet of the Georgian Academy of Sciences. **103**, 285-287 (1981). (Russian).

4. R.W.N. Kinnear, S.J. Campbell, and D.H. Chaplin. Phys. Lett. A. **76** (3–4), 311–314 (1980).

5. A.E. Reingardt, V.I. Tsifrinovich, O.V. Novoselov, and V.K. Mal'tsev. Fiz. Tverd. Tela **25** (10), 3163–3164 (1983). (Russian).

6. I.G. Kiliptari. Phys. Solid State **50** (11), 2133–2140 (2008) *[Fiz. Tverd. Tela 34 (5), 1418–1424 (1992).*

7. L.N. Shakhmuratova, D. K. Fowler, and D. Chaplin. J. Magn. Magn. Mater. **177**–**181**, 1476–1477 (1998).

8. V.S. Kuz'min, V.M. Kolesenko, and E.P. Borbotko. Phys. Solid State 50 (11), 2133–2140 (2008) *[Fiz. Tverd. Tela **50** (11), 2043–2049 (2008)]*.

9. A.L. Bloom. Phys. Rev. **98**(4), 1105-1111 (1955).

10. A.M. Akhalkatsi, T.A. Gavasheli, T.O. Gegechkori, G.I. Mamniashvili, Z.G. Shermadini, and W. G. Clark. J. Appl. Phys. **105**, 07D303 (2009).

11. A.M. Akhalkatsi and G.I. Mamniashvili. Phys. Met. Metallogr. 86 (5), 461–463 (1998*) [Fiz. Met. Metalloved. 86 (5), 64–68 (1998)]*.

12. A.M. Akhalkatsi, M.D. Zviadadze, G.I. Mamniashvili, N.M. Sozashvili, A.N. Pogorelyi, and O.M. Kuz'mak. Phys. Met. Metallogr., **98** (3), 252–260 (2004) *[Fiz. Met. Metalloved. 98 (3), 23–31 (2004)]*.

13. M.D. Zviadadze, R.L. Lepsveridze, G.I. Mamniashvili and A.M. Akhalkatsi . Phys. Met. Metallogr., **111**, N 6, 547-553 (2011).

14. M.D. Zviadadze, G.I. Mamniashvili, T.O. Gegechkori, A.M. Akhalkatsi, and T.A. Gavasheli. E-print arXiv:1204.0686v1 [cond-mat.mtvl-sci] (2012).

15. M.I. Kurkin and E.A. Turov. "NMR in magnetically ordered substances and its application". Moscow, Nauka, 244 p. (1990) (Russian).

16. H.C. Torrey. Phys. Rev., **76,** 1059-1068 (1949).

17. T. P. Das and D. K. Roy. Phys. Rev. **98**, 525–531 (1955).

18. M.D. Zviadadze**,** N.D. Chachava, A.G. Kvirikadze, A.K. Pokleba. E-print arXiv: 1006.1143 (2010).

19. A.M. Akhalkatsi, G.I. Mamniashvili, S. Ben-Ezra. Phys. Lett. A **291**, 34-38 (2001).